\documentclass[sigconf,natbib=true,screen]{acmart}
\usepackage{enumerate}
\usepackage{enumitem}
\usepackage{multirow}
\usepackage{bbding}
\usepackage{pifont}
\usepackage{wasysym}

\usepackage{amssymb}
\usepackage{bm}
\usepackage{graphicx}
\usepackage{amsmath}
\usepackage{float}
\usepackage{color}
\usepackage{subfigure}
\usepackage[normalem]{ulem}
\usepackage{colortbl}
\usepackage{makecell}
\usepackage{float}
\usepackage{booktabs}
\usepackage[noend]{algpseudocode}
\usepackage{algorithmicx,algorithm}
\usepackage{graphicx}
\usepackage{balance}
\usepackage{microtype}

\makeatletter
\newcommand*\bigcdot{\mathpalette\bigcdot@{.5}}
\newcommand*\bigcdot@[2]{\mathbin{\vcenter{\hbox{\scalebox{#2}{$\m@th#1$}}}}}
\makeatother

\useunder{\uline}{\ul}{}

\acmConference[ESEC/FSE '23]{Proceedings of the 31st ACM Joint European Software Engineering Conference and Symposium on the Foundations of Software Engineering}{December 3--9, 2023}{San Francisco, CA, USA}
\acmBooktitle{Proceedings of the 31st ACM Joint European Software Engineering Conference and Symposium on the Foundations of Software Engineering (ESEC/FSE '23), December 3--9, 2023, San Francisco, CA, USA}
\received{2023-02-02}
\received[accepted]{2023-07-27}

\begin{document}

\title{Pre-training Code Representation with Semantic Flow Graph for Effective Bug Localization}

\author{Yali Du}
\affiliation{%
  \institution{Shandong University}
  \city{}
  \state{}
  \country{}}
\email{duyali2000@gmail.com}

\author{Zhongxing Yu}
\authornote{Zhongxing Yu is the corresponding author.}
\affiliation{%
  \institution{Shandong University}
  \city{}
  \state{}
  \country{}}
\email{zhongxing.yu@sdu.edu.cn}

\begin{abstract}
Enlightened by the big success of pre-training in natural language processing, 
pre-trained models for programming languages have been widely used to promote code intelligence in recent years. In particular, BERT has been used for bug localization tasks and impressive results have been obtained. However, these BERT-based bug localization techniques suffer from two issues. First, the pre-trained BERT model on source code does not adequately capture the deep semantics of program code. Second, the overall bug localization models neglect the necessity of large-scale negative samples in contrastive learning for representations of changesets and ignore the lexical similarity between bug reports and changesets during similarity estimation. We address these two issues by 1) proposing a novel directed, multiple-label code graph representation named Semantic Flow Graph (SFG), which compactly and adequately captures code semantics, 2) designing and training SemanticCodeBERT based on SFG, and 3) designing a novel Hierarchical Momentum Contrastive Bug Localization technique (HMCBL). Evaluation results show that our method achieves state-of-the-art performance in bug localization.  

\end{abstract}

\ccsdesc[500]{Software and its engineering~Software testing and debugging}
\ccsdesc[500]{Software and its engineering~Maintaining software}


\keywords{bug localization, semantic flow graph, type, computation role, pre-trained model, contrastive learning}

\maketitle

\section{Introduction}
While modern software engineering recognizes a broad range of methods (e.g., model checking, symbolic execution, type checking) for helping ensure that the software meets the specification of its desirable behavior, the software (even deployed ones) is still unfortunately plagued with heterogeneous bugs for reasons such as programming errors made by developers and immature development process. The process of resolving the resultant bugs termed debugging, is an indispensable yet frustrating activity that can easily account for a significant part of software development and maintenance costs~\cite{debugging}. To tackle the ever-growing high costs involved in debugging, a variety of automatic techniques have been proposed as debugging aids for developers over the past decades \cite{yuemse}. In particular, numerous methods have been developed to facilitate fault localization, which aims to identify the exact locations of program bugs and is one of the most expensive, tedious, and time-consuming activities in debugging~\cite{wong2016survey,multiple-fault}. 

The literature on fault localization is rich and is abundant with methods stemming from ideas that originate from several different disciplines, notably including statistical analysis~\cite{jones2005empirical,liu2005sober,guifault,yuguifl}, program transformation~\cite{papadakis2015metallaxis}, information retrieval~\cite{saha2013improving,improving}. Among them, information retrieval-based methods typically proceed by establishing the relevance between bug reports and related software artifacts on the ground of information retrieval techniques, and this category of methods is appealing as it is amenable to the mainstream development practice which features continuous integration (CI), versioning with Git, and collaboration within platforms like GitHub~\cite{urli2018design}. In line with existing literature, information retrieval-based fault localization hereafter is simply referred to as bug localization. 

The matched software artifact at the early phase of bug localization research focuses on code elements such as classes and methods~\cite{shivaji2012reducing,kim2013should}, but recent years have witnessed a growing interest in changesets~\cite{bugdataset,wu2018changelocator,ciborowska2022online,improving}. The key advantage of changesets is that they contain simultaneously changed parts of the code that are related, facilitating bug fixing. With regard to information retrieval techniques, the major shift is that the dominating techniques have changed from Vector Space Model (VSM) to deep learning techniques, both for code elements and changesets. To precisely locate the bug, bug localization techniques essentially need to accurately relate the natural language used to describe the bug (in the bug report) and identifier naming practices adopted by developers (in the software artifacts). However, it is quite common that there exists a significant lexical gap between them, and consequently, the retrieval quality of bug localization techniques is not always satisfactory~\cite{8466000}. To overcome the issue, bug localization techniques necessarily need to go beyond exact term matching and establish the semantic relatedness between bug reports and software artifacts.

Given that deep learning architectures are capable of leveraging contextual information and have achieved impressive progress in natural language processing, a number of bug localization techniques based on the neural network have been proposed in recent years~\cite{dnnloc,ciborowska2022fast,tranpcnn,9566596,bl13,ZHU2022108741,mahajan2022design,bug2commit,xie2022universal}. In particular, state-of-the-art transformer-based architecture BERT~\cite{bertpaper} (bidirectional encoder representation from the transformer) has been widely employed~\cite{ciborowska2022fast,icse21bert}. Based on the naturalness hypothesis which states that ``\emph{software corpora have similar statistical properties to natural language corpora}''~\cite{Naturalness}, these BERT-based techniques first pre-train a BERT model on a massive corpus of source code using certain pre-training tasks such as masked language modeling, and then fine-tune the trained BERT model for bug localization task. Experimental evaluations have shown that reasonable accuracy improvements can be obtained by these BERT-based techniques. 

Despite the progress made, one drawback of these BERT-based techniques is that the pre-trained BERT model on source code does not adequately capture the deep semantics of program code. Unlike natural language, the programming language has a formal structure, which provides important code semantics that is unambiguous in general~\cite{surveycode}. However, the existing pre-trained BERT model either totally ignores the code structure by treating code snippet as a sequence of tokens same as natural language or considers only the shallow structure of the code by using graph code representations such as data flow graph~\cite{graphcodebert}. Consequently, the formal code structure has not been fully exploited, resulting in an under-optimal BERT model. To overcome this issue, we in this paper present a novel code graph representation termed Semantic Flow Graph (SFG), which compactly and adequately captures code semantics. SFG is a directed, multiple-label graph that captures not only the data flow and control flow between program elements but also the type of program element and the specific role that a certain program element plays in computation. On the ground of SFG, we further propose SemanticCodeBERT, a pre-training model with BERT-like architecture to learn code representation that considers deep code structure. SemanticCodeBERT features novel pre-training tasks besides the ordinary masked language modeling task.

In addition, the overall models of existing BERT-based bug localization techniques ignore several points which are beneficial for further improving performance. 
First, the batch size is typically limited to save model space because of the huge scale of BERT parameters, and the number of negative samples coupled to batch size is thus limited. A variety of existed methods~\cite{moco,chen2020simple,wu2018unsupervised,mocov2,emoco,con1,con2,con3,con4,con5,con6,con7,con8,con9,con10,DuW0L0N23} emphasizes the necessity of large-scale negative samples in contrastive representation learning. In the bug localization context, it implies the importance of considering the large-scale negative sample interactions for representation learning of bug reports and changesets. Nevertheless, existing techniques like Ciborowska \textit{et. al.}~\cite{ciborowska2022fast} only select one irrelevant changeset in training as the negative sample for a bug report, which causes inefficient mining of negative samples and poor representation of the programming language.
To alleviate this issue, we propose to use a memory bank~\cite{wu2018unsupervised} to store rich changesets obtained from different batches for later contrast. In particular, due to the constant parameter update by back-propagation, we utilize the momentum contrastive method~\cite{moco} to account for the inconsistency of negative vectors obtained by different models (in different mini-batches). 
Second, existing BERT-based bug localization techniques only account for the semantic level similarity between bug reports and changesets, totally ignoring the lexical similarity (\emph{e.g.}, same identifier) which is also of vital importance for retrieval if exists. To alleviate this issue, we propose to use a hierarchical contrastive loss to leverage similarities at different levels. On the whole, we design a novel Hierarchical Momentum Contrastive Bug Localization (HMCBL) technique to address the two limitations.

We implement the analyzer for obtaining SFG for Java code and use the Java corpus (including 450,000 functions) of the CodeSearchNet dataset~\cite{codesearchnet} to pre-train SemanticCodeBERT. On top of SemanticCodeBERT, we apply the hierarchical momentum contrastive method to facilitate the retrieval of bug-inducing changesets given a bug report on the widely used dataset established in~\cite{bugdataset}, which includes six Java projects. Results show that we achieve state-of-the-art performance on bug localization. Ablation studies justify that the newly designed SFG improves the BERT model and the new bug localization architecture is better than the existing ones. 

Our contributions can be summarized as follows: 
\begin{itemize}[leftmargin=*]
\item We present a novel directed, multiple-label code graph representation termed Semantic Flow Graph (SFG), which compactly and adequately captures code semantics. 
\item We employ SFG to train SemanticCodeBERT, which can be applied to obtain code representations for various code-related downstream tasks.
\item We design a novel Hierarchical Momentum Contrastive Bug Localization technique (HMCBL), which overcomes two important issues of existing techniques. 
\item We conduct a large-scale experimental evaluation, and the results show that our method outperforms state-of-the-art techniques in bug localization performance.  
\end{itemize}

\section{Related works}
This section reviews work closely related to this paper.
Bug localization techniques proceed by making a query about the relevance between bug reports and related software artifacts on top of information retrieval techniques. The investigated software artifacts can be majorly divided into two categories: code elements such as classes and methods~\cite{shivaji2012reducing,kim2013should,saha2013improving,bl5,bl6,bl7,bl8,bl9,bl11,bl12,bl13,MaL1,MaL2,HuoLZ20,li2021laprob} and changesets~\cite{bugdataset,wu2018changelocator,ciborowska2022online,improving}. Given changesets contain simultaneously changed parts of the code that are related and can thus facilitate bug fixing, the use of changesets is gradually dominating.

With regard to information retrieval techniques, the
Vector Space Model (VSM) is widely used for its simplicity and effectiveness, especially in the early phase of bug localization research. For instance, BugLocator~\cite{buglocator} makes use of the revised Vector Space Model (rVSM) to establish the textual similarity between the bug report and the source code and then ranks all source code files based on the calculated similarity. For another example, Locus~\cite{bugdataset} represents one of the earliest works on changeset-based bug localization, and it proceeds by matching bug reports to hunks. 

As VSM basically performs exact term matching, the effectiveness will be compromised in the common case where there exists a significant lexical gap between the descriptions in the bug report and naming practices adopted by developers in the software artifacts. To overcome this issue, bug localization techniques essentially need to establish the semantic relatedness between bug reports and software artifacts. Given the impressive progress in leveraging contextual information by deep learning architectures in natural language processing, deep neural networks have been widely used by researchers to learn representations for bug localization in recent years~\cite{cooba, dnnloc, tranpcnn, bug2commit}. 
For instance, Huo \textit{et. al.}~\cite{tranpcnn} present the Deep Transfer Bug Localization task, and propose the TRANP-CNN as the first solution for the cold-start problem which combines cross-project transfer learning and convolutional neural networks for file-level bug localization. 
Zhu \textit{et. al.}~\cite{cooba} focus on transferring knowledge (while filtering out irrelevant noise) from the source project to the target project, and propose the COOBA to leverage adversarial transfer learning for cross-project bug localization. 
Murali \textit{et. al.}~\cite{bug2commit} propose Bug2Commit, which is an unsupervised model leveraging multiple dimensions of data associated with bug reports and commits.

In particular, enlightened by the impressive achievements made by BERT in natural language
processing, BERT has been used for bug localization tasks. 
Lin \textit{et. al.}~\cite{icse21bert} study the tradeoffs between different BERT architectures for the purpose of changeset retrieval. 
Based on the Colbert developed by Khattab \textit{et. al.}~\cite{colbert}, Ciborowska \textit{et. al.}~\cite{ciborowska2022fast} propose the FBL-BERT model towards changeset-based bug localization. Evaluation results show that FBL-BERT can speed up the retrieval and several design decisions have also been explored, including granularities of input changesets and the utilization of special tokens for capturing changesets' semantic representation. 
While impressive retrieval results of changesets have been achieved, the Colbert used by FBL-BERT does not adequately capture the deep semantics of program code and the overall models of FBL-BERT suffer from two important limitations as described in Section 1 (Introduction).

Furthermore, inspired by the success of pre-training models in natural language processing, a number of pre-trained models for programming languages have been proposed to promote the development of code representation (which is vital for a variety of code-based tasks in the field of SE). 
For instance, 
CodeBERT is a pre-trained model proposed by Feng \textit{et. al.}~\cite{codebert}, which provides generic representations for natural and programming language downstream applications.
GraphCodeBERT~\cite{graphcodebert} imports structural information to enhance the code representation by adding the data flow graph as an auxiliary of input tokens and improves the performance of code representation compared to CodeBERT. 
Kanade \textit{et. al.}~\cite{kanade} propose CuBERT, which is pre-trained on a massive Python source corpus with two pre-training tasks of Masked Language Modeling (MLM) and Next Sentence Prediction (NSP).
Buratti \textit{et. al.}~\cite{cbert} propose C-BERT, a transformer-based language model that is pre-trained on the C language corpus for code analysis tasks.
Xue \textit{et. al.}~\cite{treebert} propose TreeBERT, which proposes a hybrid target for AST to learn syntactic and semantic knowledge with tree-masked language modeling (TMLM) and node order prediction (NOP) pre-training tasks.
More recently, the UniXcoder~\cite{unixcoder} is proposed to leverage cross-modal information like Abstract Syntax Tree and comments written in natural language to enhance code representation. 
While these pre-trained models on source code have made progress towards code representation, one drawback of them is that they
not adequately capture the deep semantics of program code as they either treat code snippets as token sequences or consider
only shallow code structure by using graph code representations such as data flow graph. Hence, we give a novel code graph representation termed Semantic Flow Graph (SFG) to more compactly and adequately capture code semantics in this paper. On top of SFG, we further design and train SemanticCodeBERT with novel pre-training tasks. 

\vspace{2mm}
\begin{figure*}
    \centering
    \includegraphics[width=0.95\textwidth]{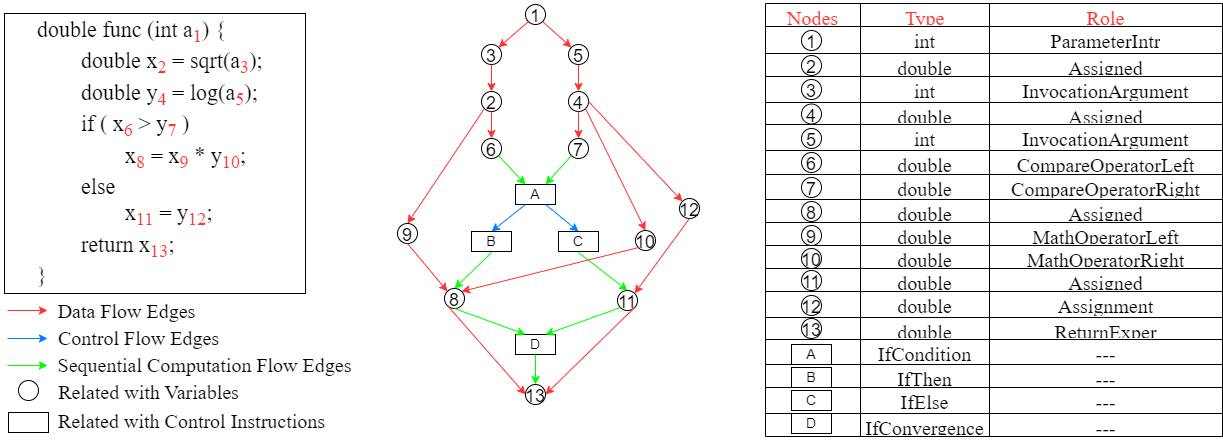}
    \caption{An example of the semantic flow graph.}
    \label{SFG}
\end{figure*}

\section{Semantic Flow Graph}\label{Semantic Flow Graph}
This section introduces the Semantic Flow Graph (SFG), a novel code graph representation designed for compactly representing deep code semantics. On top of the naturalness hypothesis ``\emph{Software is a form of human communication, software corpora have similar statistical properties to natural language corpora, and these properties can be exploited to build better software engineering tools}''~\cite{Naturalness}, recent years have witnessed many innovations in using machine learning (particularly deep learning) techniques to help make the software more reliable and maintainable. To achieve successful learning, one important ingredient lies in suitable code representation. The representation, on the one hand, should capture enough code semantics, and on the other hand, should be learnable across code written by different developers or even different programming languages~\cite{surveycode}. 

There are majorly three categories of code representation ways within the literature: token-based ways that represent code as a sequence of tokens~\cite{Miltiadis2, Jinghui,8091247,Gupta}, syntactic-based ways that represent code as trees~\cite{Miltiadis1,Yuhan,Lili,Michael,Zilberstein,code2vec, yutse}, and semantic-based ways that represent code as graph structures~\cite{Kwonsoo,allamanis2018learning,Kremenek,david2020neural,Jakobovits,graphcodebert,DependencyGraph,Yaqin}. For token-based representation, while its simplicity facilities learning, the representation ignores the structural nature of code and thus captures quite limited semantics. For syntactic-based representation, despite the tree representation in principle can contain rich semantic information, the learnability is unfortunately confined as the tree typically has an unusually deep hierarchy and there, in general, will involve significant refinement efforts of the raw tree representation to enable successful learning in practice. Semantic-based representation aims to encode semantics in a way that facilitates learning, and a variety of graphs have been employed for code model learning, including for example data flow graph~\cite{graphcodebert,Kwonsoo}, control flow graph~\cite{david2020neural}, program dependence graph~\cite{DependencyGraph}, contextual flow graph~\cite{Jakobovits}. 

While these graph-based representations have facilitated the learning of code semantics embodied in data dependency and control dependency, certain other code semantics are overlooked. In particular, the information of \emph{what} kinds of program elements are related by data dependency or control dependency and through \emph{which} operations they are related to is neglected. We argue that this information is crucial for accurately learning code semantics. For instance, given a code snippet ``\texttt{a = m (b, c)}'' where \texttt{a}, \texttt{b}, and \texttt{c} are \texttt{Boolean}, \texttt{Integer}, and User-defined type variables respectively, and \texttt{m} is a certain function call, there will be two data flow edges $\texttt{b} \rightarrow \texttt{a}$ and $\texttt{c} \rightarrow \texttt{a}$ considering the data flow graph, and the code snippet will read ``\emph{the values of two variables have flown into another variable}''. Under this circumstance, as the corresponding data flow graph coincides, the meaning of the code snippet has no difference with a variety of other code snippets such as ``\texttt{a = (b \&\& m (c))}'' where \texttt{a}, \texttt{b}, and \texttt{c} are \texttt{Boolean}, \texttt{Boolean}, and arbitrary type variables respectively, and \texttt{m} is a certain function call that returns a boolean value. But if the additional information of \emph{what} kinds of program elements and \emph{which} operations are taken into account, the code snippet ``\texttt{a = m (b, c)}'' will read ``\emph{the value of an Integer type variable and the value of a User-defined type variable have flown into another Boolean type variable through a function call}'', which is more precise. To compactly integrate these two pieces of information into graphs, we design a novel directed, multiple-label code graph representation termed Semantic Flow Graph (SFG).

\vspace{1.0mm}
\noindent 
\textbf{Definition 3.1.} \textbf{(Semantic Flow Graph)}. The Semantic Flow Graph (SFG) for a code snippet is a tuple $<N, E, T, R>$ where \emph{N} is a set of nodes, \emph{E} is a set of directed edges between nodes in \emph{N}, and \emph{T} and \emph{R} are mappings from nodes to their types and their roles in computation respectively. 
\vspace{1.0mm}

A number of points deserve comment. First, the node set \emph{N} can be further divided into node sets $\emph{N}_{V}$ and $\emph{N}_{C}$, which contain nodes corresponding to variables and control instructions in the code respectively. While the variable has a one-to-one mapping with a certain node from $\emph{N}_{V}$, there may be one or multiple nodes in $\emph{N}_{C}$ for a certain control instruction. Essentially, if a control instruction has an associated condition and \emph{n} different branches (i.e., straight-line code blocks) to go depending on the condition evaluation result, there will be a node in $\emph{N}_{C}$ for the condition, a node in $\emph{N}_{C}$ for the convergence of the different branches, and \emph{n} different nodes in $\emph{N}_{C}$ for the \emph{n} branches respectively. 

Second, a directed edge $\emph{n}_{a} \rightarrow \emph{n}_{b}$ ($\emph{n}_{a} \in \emph{N}, \emph{n}_{b} \in \emph{N}$) in \emph{E} can be of 3 kinds. The first kind $\emph{E}_{D}$ represents a data flow between two variables if $\emph{n}_{a} \in \emph{N}_{V} \land \emph{n}_{b} \in \emph{N}_{V}$ holds, the second kind $\emph{E}_{C}$ embodies the control flow between two straight-line basic blocks if $\emph{n}_{a} \in \emph{N}_{C} \land \emph{n}_{b} \in \emph{N}_{C}$ holds, and finally the third kind $\emph{E}_{S}$ denotes the natural sequential computation flow inside or between basic blocks in case $\emph{n}_{a} \in \emph{N}_{V} \land \emph{n}_{b} \in \emph{N}_{C}$ or $\emph{n}_{a} \in \emph{N}_{C} \land \emph{n}_{b} \in \emph{N}_{V}$ holds. In particular, the edge set \emph{E} is established as follows: 

\vspace{1.0mm}
\noindent \emph {(1) Establish} $\emph{E}_{D}$ \emph {among nodes from set} $\emph{N}_{V}$ \emph {according to Intra-block and Inter-block data dependencies between variables.}

\vspace{1.0mm}
\noindent \emph {(2) Establish} $\emph{E}_{C}$ \emph {among nodes from set} $\emph{N}_{C}$ \emph {according to the specific control flow of the control instruction.}

\vspace{1.0mm}
\noindent \emph {(3) Establish} $\emph{E}_{S}$ \emph {following these rules: there will be an edge} $\emph{n}_{a} \rightarrow \emph{n}_{b}$ \emph {(i) if} $\emph{n}_{b} \in \emph{N}_{C}$ \emph {is for the control instruction condition and }$\emph{n}_{a} \in \emph{N}_{V}$ \emph {is for a certain variable involved in the condition; (ii) if}  $\emph{n}_{a} \in \emph{N}_{C}$ \emph {is for the control instruction branch and }$\emph{n}_{b} \in \emph{N}_{V}$ \emph {is for the left-most variable of the first statement inside the branch; (iii) if} $\emph{n}_{a} \in \emph{N}_{V}$\emph { is for the left-most variable of the last statement inside a control instruction branch and }$\emph{n}_{b} \in \emph{N}_{C}$ \emph {is for the control instruction convergence; (IV) if} $\emph{n}_{a} \in \emph{N}_{C}$ \emph {is for the control instruction convergence and }$\emph{n}_{b} \in \emph{N}_{V}$ \emph {is for the left-most variable of the first statement inside the basic block directly following the control instruction.}
\vspace{1.0mm} 

Third, mapping \emph{T} maps each node in \emph{N} to its type, encoding the needed information of ``\emph{what} kinds of program elements are related''. For each node in $\emph{N}_{V}$, \emph{T} maps it to the corresponding type of the variable. For each node in $\emph{N}_{C}$, \emph{T} maps the node to the specific part of the control instruction it refers to. Take the control instruction \texttt{If-Then-Else} as an example, \emph{T} maps the associated 4 nodes in $\emph{N}_{C}$ for it to type \texttt{IfCondition}, \texttt{IfThen}, \texttt{IfElse}, and \texttt{IfCONVERGE} respectively.

Finally, mapping \emph{R} maps each node in $\emph{N}_{V}$ to its role in the computation, encoding the needed information of ``through \emph{which} operations program elements are related''. Basically, \emph{R} considers the associated operation and control structure for the variable to determine its computation role. From an implementation perspective, for each node in $\emph{N}_{V}$, \emph{R} checks the direct parent of the corresponding variable in the abstract syntax tree (AST) and the position relationship between it and the direct parent to establish the role. For instance, given a code snippet, ``\texttt{a = b}'' where \texttt{a} and \texttt{b} are variables, \emph{R} maps the roles of \texttt{a} and \texttt{b} to \texttt{Assigned} and \texttt{Assignement} respectively. For another example, given a code snippet ``\texttt{a.m(b)}'' where \texttt{a} and \texttt{b} are variables, and \texttt{m} is a certain function call, \emph{R} maps the roles of \texttt{a} and \texttt{b} to \texttt{InvocationTarget} and \texttt{InvocationArgument} respectively. For nodes in $\emph{N}_{C}$, we do not consider their roles as they are implicit in their types. 

Note it is difficult to simply augment classical 
graph program representations with information of type and computation role. Existing representations like program dependence graph typically work at the statement granularity (i.e., each graph node represents a statement), making it hard to encode detailed type and computation role information of multiple program elements in the statement. The proposed SFG works at a finer granularity with two types of nodes that have a one-to-one mapping with program variables and a one-to-one (or many-to-one) mapping with program control ingredients respectively. This kind of node representation is proposed for two reasons. On the one hand, it is convenient to analyze the types and computation roles of variables (through how they are connected with other program elements) and program control ingredients. On the other hand, data flow and control flow information are established respectively by analyzing variable uses and program control ingredients. With SFG (built on such a node representation), data flow and control flow can be encoded through the edges between nodes, and the type and computation role information can be encoded through node labels. SFG does not have nodes for additional program elements (like invocation etc.), thus it is compact but contains adequate semantic information.

Overall, SFG is a directed, multiple-label graph that captures not only the data flow and control flow between program elements, but also the type of program element and the specific role that a certain program element plays in computation. Moreover, SFG represents this information in a compact way, facilitating learning across programs. 

\vspace{1.0mm}
\noindent 
\textbf{Example 3.1.} \emph{Figure~\ref{SFG} gives an example of a Semantic Flow Graph for a simple method.} 
\vspace{1.0mm}

\noindent 
\textbf{Implementation}: We fully implement an analyzer to get Semantic Flow Graph (SFG) for a Java method on top of Spoon~\cite{spoon}, which is an open-source library to analyze, rewrite, transform, and transpile Java source code. Our analyzer supports modern Java versions up to Java 16. For nodes in $\emph{N}_{V}$, the analyzer considers different kinds of primitive types and common JDK types, and a special type named \texttt{user-defined type}. In total, the analyzer considers 20 types for nodes in $\emph{N}_{V}$. For nodes in $\emph{N}_{C}$, the analyzer takes all the control instruction kinds (up to Java 16) into account and considers 35 types in total. With regard to role, the analyzer considers 43 different roles in total for nodes in $\emph{N}_{V}$.

\vspace{2mm}
\begin{figure}
    \centering
    \includegraphics[width=0.45\textwidth]{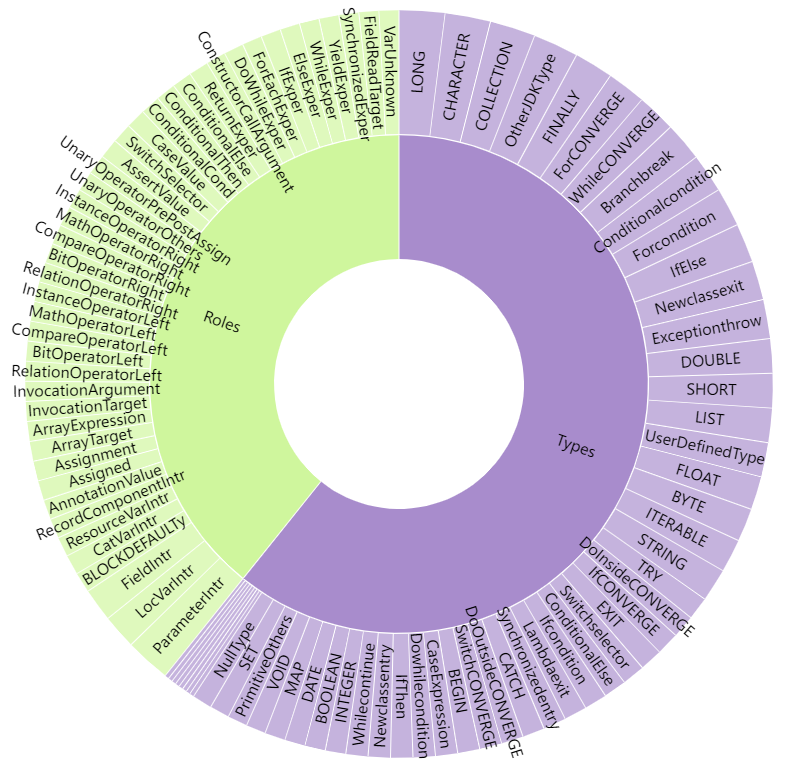}
    \caption{All defined types and roles.}
    \label{attribute}
\end{figure}

\section{SemanticCodeBERT}\label{SemanticCodeBERT}
In this section, we describe first the architecture of SemanticCodeBERT (shown in Figure~\ref{modelarchitecture}), then the graph-guided masked attention based on the semantic flow graph, and finally the pre-training tasks. 
Overall, the SemanticCodeBERT network architecture adapts the architecture of GraphCodeBERT for the proposed novel SFG program representation and SemanticCodeBERT also features tailored pre-training tasks for the SFG representation.

\subsection{Model Architecture}
The SemanticCodeBERT follows BERT (Bidirectional Encoder Representation from Transformers) (Devlin \textit{et. al.,}~\cite{bertpaper}) as the backbone.

\vspace{0.8mm}
\noindent\textbf{Comment Input Sequence}:
We import comments as a supplement for the model to understand the semantic information of programming code. $[CLS]$ is the special classification token at the beginning of the comment sequence $W$.

\vspace{0.8mm}
\noindent\textbf{Source Code Input Sequence}:
We cleanse the source code and remove erroneous characters, and add the special classification token $[SEP]$ at the end of the source code and input sequence. To represent the start-of-code, we import a pre-appended token $[C]$ to split the comment and source code. The source code sequence can be represented as $S$.

\vspace{0.8mm}
\noindent\textbf{Node Input Sequence}:
With the procedure discussed in Section \ref{Semantic Flow Graph},  we generate a semantic flow graph (SFG) for each code snippet. At the beginning of the node list $N$, a pre-appended token $[N]$ is added to represent the start-of-node.

\vspace{0.8mm}
\noindent\textbf{Type Input Sequence}:
To answer the question of "\textit{what kinds of program elements are related}", we have identified $55$ possible types for the code element. $T=\{t_1,\ldots, t_{55}\}$ represents the set of all $55$ possible types, and $[T]$ is pre-appended as the start-of-type. The complete list of types is shown in Figure \ref{attribute}.

\vspace{0.8mm}
\noindent\textbf{Role Input Sequence}:
To answer the question of "\textit{through which operations program elements are related}", we have defined $43$ roles to mark the role of each program element in the computation, taking into account the associated operation and control structure. $R=\{r_1,\ldots, r_{43}\}$ is the set of all $43$ possible roles, and the pre-appended token $[R]$ represents the start-of-role. The complete list of roles is shown in Figure \ref{attribute}.

\begin{figure*}
\centering
\includegraphics[width=\textwidth]{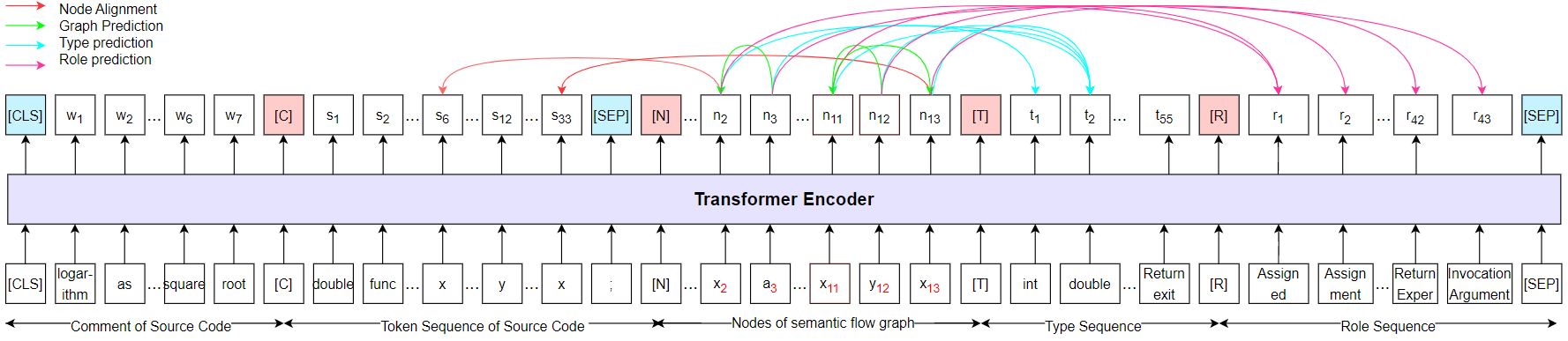}
\caption{The SemanticCodeBERT takes to comment, source code, nodes of SFG, types, and roles as the input, and is pre-trained by standard masked language modeling~\cite{bertpaper}, node alignment (marked with red lines), graph prediction (marked with green lines), type prediction (marked with blue lines) and role prediction (marked with purple lines).}
\label{modelarchitecture}
\end{figure*}

As intuitively shown in Figure \ref{modelarchitecture}, we concatenate the comment, source code, nodes, types, and roles as the input sequence: 
\begin{equation}
X = Concat[[CLS], W, [C], S, [SEP], [N], N, [T], T, [R], R, [SEP]]\text{.}
\end{equation}

\vspace{2mm}

\subsection{Masked Attention}
We resort to the graph-guided masked attention function described in ~\cite{graphcodebert} to filter irrelevant signals in Transformer. 
\begin{itemize}[leftmargin=*]
\item The set $E^1$ indicates the alignment relation between $S$ and $N$, where $(s_i,n_j)/(n_j,s_i)\in E^1$ if the node ${n_j}$ is identified from the source code token $s_i$. 
\item The set $E^2$ indicates the dependency relation in $N$, where $(n_i,n_j)\in E^2$ if there is a direct edge from the node $n_i$ to the node $n_j$.
\item The set $E^3$ incorporates the type information of the nodes, where $(n_i,t_j)\in E^3$ if the type of the node $n_i$ is $t_j$.
\item The set $E^4$ incorporates the role information of the nodes, where $(n_i,r_j)\in E^4$ if the role of the node $n_i$ is $r_j$.
\end{itemize}
 
The masked attention matrix is formulated as $M$:
\begin{equation}
    M_{ij} = \left\{
            \begin{array}{rcl}
            0 & & x_i \in {[CLS], [SEP]}\text{;} \\
            & & w_i,s_j \in W \cup S\text{;} \\
            & & (s_i,n_j)/(n_j,s_i) \in E^{1}\text{;}\\
            & & (n_i,n_j) \in E^2\text{;}\\
            & & (n_i,t_j) \in E^3\text{;}\\
            & & (n_i,r_j) \in E^4\text{;} \\
            -\infty& & otherwise\text{.}
            \end{array}
            \right. 
\end{equation}
Specifically, the masked attention function blocks the transmission of unrelated tokens by setting the attention score to an infinitely negative value.

\subsection{Pre-training Tasks}
The pre-training tasks of SemanticCodeBERT are described in this section. Besides masked language modeling, node alignment, and edge prediction pre-training tasks proposed by Guo \textit{et. al.,}~\cite{graphcodebert}, we define two novel pre-training tasks--type and role prediction. These two novel pre-training tasks represent the first attempt to leverage the attribute information of nodes for learning code representation.

\vspace{0.8mm}
\noindent\textbf{Masked Language Modeling}:
The masked language modeling pre-training task is proposed by Devlin \textit{et. al.,}~\cite{bertpaper}. We replace 15\% of the source code with [MASK] 80\% of the time, a random token 10\% of the time or itself 10\% of the time. The comment context contributes to inferring the masked code tokens~\cite{graphcodebert}. 

\vspace{0.8mm}
\noindent\textbf{Node Alignment}:
The motivation of node alignment is to align representation between source code and nodes of semantic flow graph~\cite{graphcodebert}. We randomly mask 20\% edges between the source code and nodes, and then predict where the nodes are identified from (i.e., predict these masked edges $E^1_{mask}$). As shown in the Figure~\ref{modelarchitecture}, the model should distinguish that $n_2$ comes from $s_6$ and $n_{13}$ comes from $s_{33}$. We formulate the loss function as Equation~\ref{nodealign}. Let $E^1$ be $S \times N$, $\delta(e_{ij}\in E^1_{mask})$ is one if $(s_i,n_j)\in E^1$, and zero otherwise. $p_{e_{ij}}$ is the probability of the edge from $i$-th code token to $j$-th node, which is calculated by dot product following a sigmoid function using the representations of $s_i$ and $n_j$ outputted from SemanticCodeBERT. 
\begin{equation}
\begin{aligned}
\mathcal{L}_{NA} = - \sum_{e_{ij} \in E^1_{mask}}[\delta(e_{ij})log p_{e_{ij}} + \\
(1-\delta(e_{ij}))log (1-p_{e_{ij}})]\text{.}
\label{nodealign}
\end{aligned}
\end{equation}
\\
\textbf{Edge Prediction}:
The motivation of edge prediction is to encourage the model to learn structural relationships from semantic flow graphs for better programming code representation. Like node alignment, we randomly mask 20\% edges between nodes in the mask matrix, encouraging the model to predict these masked edges $E^2_{mask}$ (\textit{e.g.,} the edges ($n_3$, $n_2$) and ($n_{12}$, $n_{11}$)). 
We formulate the loss function as Equation~\ref{graphpred}. Let $E^2$ be $N \times N$, $\delta(e_{ij}\in E^2_{mask})$ is one if $(n_i,n_j)\in E^2$, and zero otherwise. $p_{e_{ij}}$ is the probability of the edge from $i$-th node to $j$-th node.
\begin{equation}
\begin{aligned}
\mathcal{L}_{GP} = - \sum_{e_{ij} \in E^2_{mask}}[\delta(e_{ij})log p_{e_{ij}} + \\
(1-\delta(e_{ij}))log (1-p_{e_{ij}})]\text{.}
\label{graphpred}
\end{aligned}
\end{equation}
\\
\textbf{Type Prediction}:
The motivation of type prediction is to guide the model to comprehend the types (\textit{e.g.,} ``int'', ``double'', ``IfCondition'') of nodes for better programming code representation. 
We pre-append the full set of types $T$ to the input nodes. Let $E^3$ be $ N \times T$, if the type of node $n_i$ is $t_j$ (\textit{i.e.,} $(n_i, t_j)\in E^3$), 
$\delta(e_{ij}\in E^3_{mask})$ is one, otherwise it is zero. We randomly mask 20\% edges between nodes and types and formulate the loss function as Equation~\ref{typepred}, where $E^3_{mask}$ are masked edges and $p_{e_{ij}}$ is the probability of the edge from $i$-th node to $j$-th type.
\begin{equation}
\begin{aligned}
\mathcal{L}_{TP} = - \sum_{e_{ij} \in E^3_{mask}}[\delta(e_{ij})log p_{e_{ij}} + \\
(1-\delta(e_{ij}))log (1-p_{e_{ij}})]\text{.}
\label{typepred}
\end{aligned}
\end{equation}
\\
\textbf{Role Prediction}:
``Role'' indicates the computation role of the node in the semantic flow graph (\textit{e.g.,} ``InvocationArgument'', ``Assigned'',  ``Assignment''). Role prediction can feed the model with a more informative signal to understand the correlation among different nodes. We pre-append the full set of roles $R$ to the input nodes. Let $E^4$ be $ N \times R$, if the role of node $n_i$ is $r_j$ (\textit{i.e.,} $(n_i, r_j)\in E^4$), 
$\delta(e_{ij}\in E^4_{mask})$ is one, otherwise it is zero. We randomly mask 20\% edges between nodes and roles and formulate the loss function as Equation~\ref{rolepred}, where $E^4_{mask}$ are masked edges and $p_{e_{ij}}$ is the probability of the edge from $i$-th node to $j$-th role.
\begin{equation}
\begin{aligned}
\mathcal{L}_{RP} = - \sum_{e_{ij} \in E^4_{mask}}[\delta(e_{ij})log p_{e_{ij}} + \\
(1-\delta(e_{ij}))log (1-p_{e_{ij}})]\text{.}
\label{rolepred}
\end{aligned}
\end{equation}

\vspace{2mm}
\begin{figure*}
\centering
\includegraphics[width=\textwidth]{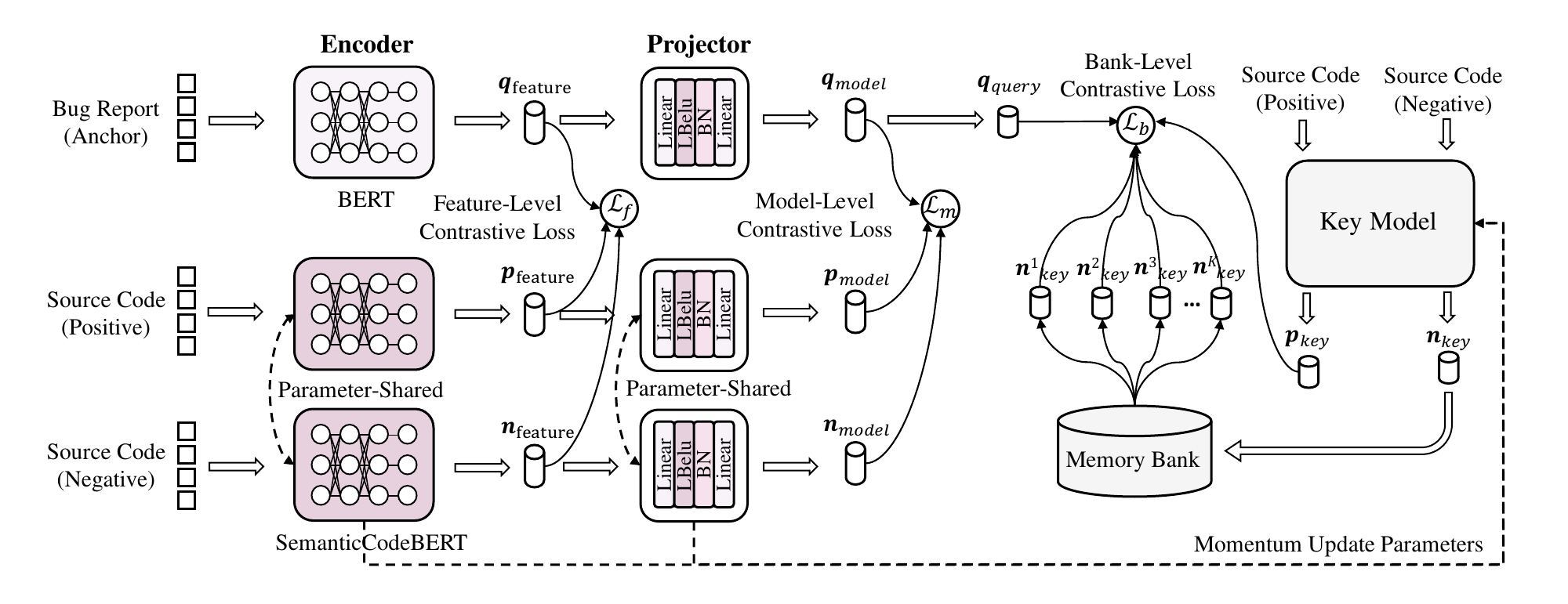}
\caption{An overview of the Hierarchical Momentum Contrastive Bug Localization technique (HMCBL).}
\label{retrievalmodel}
\end{figure*}

\section{Changeset-based bug localization}
In this section, we illustrate the utilization of the SemanticCodeBERT towards bug localization with changesets. 
The proposed bug localization model is shown in Figure~\ref{retrievalmodel}. The model aims to address the two important limitations (as described in Section 1) of the overall models of existing BERT-based bug localization techniques.

\subsection{Problem Definition}
Given a set $\mathcal{Q}=\{q_1, q_2, \dots, q_M\}$ of $M$ bug reports, the bug localization task aims to discover more relevant changesets from $\mathcal{K}=\{k_1,k_2,\dots, k_N$\}, a set including $N$ changesets. 
More specifically, for a bug report $q\in\mathcal{Q}$, a bug-inducing changeset $p\in\mathcal{K}$ and a not bug-inducing changeset $n\in\mathcal{K}$ are selected to form a triplet ($q,p,n$). All bug-inducing changesets and not bug-inducing changesets are non-overlapping.
The goal of learned similarity function $s$ is to provide a high value for $s(q,p)$ (between the anchor $q$ and the positive sample $p$) and a low value for $s(q,n)$ (between the anchor $q$ and the negative sample $n$). 
Section~\ref{Representation learning} focuses on producing accurate representations of bug reports and changesets, and Section~\ref{Similarity Estimation} describes the estimation of similarities and the loss function for training the model.

\subsection{Representation learning} \label{Representation learning}
The proposed model consists of three parts, an encoder network, projector network, and momentum update mechanism with a memory bank that stores rich representations of changesets.

\vspace{0.8mm}
\noindent
\textbf{Encoder Network}:
As mentioned before, bug reports consist of natural language descriptions and project changesets consist of programming language code. Hence, we introduce BERT~\cite{bertpaper} as the backbone to the encoder bug report as $\mathbf{q}_{feature}$, and SemanticCodeBERT as the backbone to encoder relevant changeset as $\mathbf{p}_{feature}$ and irrelevant changeset as $\mathbf{n}_{feature}$. 
\begin{equation}
    \begin{cases}
    \mathbf{q}_{feature} = \mathbf{BERT}(q_{tok})\text{,}\\
    \mathbf{p}_{feature} = \mathbf{SemanticCodeBERT}(p_{tok})\text{,}\\
    \mathbf{n}_{feature} = \mathbf{SemanticCodeBERT}(n_{tok})\text{,}
    \end{cases}
\end{equation}
where $\mathbf{BERT}$ and $\mathbf{SemanticCodeBERT}$ are the trainable parameters of BERT and SemanticCodeBERT, $q_{tok}$, $p_{tok}$, and $n_{tok}$ are the input tokens obtained by tokenizers, $\mathbf{q}_{feature}\in\mathbb{R}^{d}$, $\mathbf{p}_{feature}\in\mathbb{R}^{d}$, and $\mathbf{n}_{feature}\in\mathbb{R}^{d}$ are the refined vectors ($d$ is the dimension of the mapped spaces). 

\vspace{0.8mm}
\noindent\textbf{Projector Network}:
After the feature vectors are extracted, we use a multi-layer perception neural network as a projector to compress the vectors of bug reports and changesets into a compact shared embedding space. We replace Dropout with Batch Normalization for regularization, which can be trained with saturating nonlinearities and are more tolerant to increased training rates~\cite{batchnorm}.
\begin{equation}
    \begin{cases}
    \mathbf{q}_{model} = \mathbf{W}^2_{b}norm(\phi(\mathbf{W}^1_b\mathbf{q}_{feature}))\text{,}\\
    \mathbf{p}_{model} = \mathbf{W}^2_{c}norm(\phi(\mathbf{W}^1_c\mathbf{p}_{feature}))\text{,}\\
    \mathbf{n}_{model} = \mathbf{W}^2_{c}norm(\phi(\mathbf{W}^1_c\mathbf{n}_{feature}))\text{,}
    \end{cases}
\end{equation}
where $\mathbf{q}_{model}\in\mathbb{R}^{d'}$, $\mathbf{p}_{model}\in\mathbb{R}^{d'}$, and $\mathbf{n}_{model}\in\mathbb{R}^{d'}$ are the projected vectors ($d'$ is the dimension of the output of projector), $\mathbf{W}^{\cdot}_b$ and $\mathbf{W}^{\cdot}_c$ are the trainable weight matrices, $norm(\cdot)$ denotes batch normalization~\cite{batchnorm}, and $\phi(\cdot)$ is the $leaky\_relu$ function~\cite{leakyrelu}. 

\vspace{0.8mm}
\noindent
\textbf{Momentum Update Mechanism with Memory Bank}:
As mentioned in Section 1, it is important to consider large-scale negative samples in
contrastive learning for representations of changesets. To account for this, we use memory bank~\cite{wu2018unsupervised} to store rich changesets obtained from different batches for later contrast. In particular, 
we build the key model for encoder and projector networks of changesets based on the momentum contrastive learning mechanism proposed by He \textit{et. al.}~\cite{moco}. The parameters of the query model $\theta^q$, are updated by back-propagation, while the parameters of the key model $\theta^k$ are momentum updated as follows:
\begin{equation}
    \theta^k \gets m\theta^k + (1 - m)\theta^q\text{,}
\end{equation}
where $m \in$ $[0, 1)$ is a pre-defined momentum coefficient, which is set as $0.999$ in our experiment. As proved in the previous study~\cite{moco}, a relatively large momentum works much better than a smaller value suggesting that a slowly evolving key model is core to making use of the memory bank.
For per mini-batch, we use average pooling and enqueue the latest negative samples into the memory bank and dequeue the oldest negative samples.

\subsection{Similarity Estimation} \label{Similarity Estimation}
As mentioned before, the lexical similarity between bug reports and program changesets like the same application programming interfaces is also crucial for retrieval besides semantic similarity. In this paper, we use the hierarchical contrastive loss to leverage the lower feature-level similarity, higher model-level similarity, and broader bank-level similarity for matching the bug report with relevant changesets.
We get the positive feature-level similarity $s^{f+}$ by calculating cosine similarity between $q_{feature}$ and $p_{feature}$, the negative feature-level similarity $s^{f-}$ by calculating cosine similarity between $q_{feature}$ and $n_{feature}$, the positive model-level similarity  $s^{m+}$ by calculating cosine similarity between $q_{model}$ and $p_{model}$, and the negative model-level similarity $s^{m-}$ by calculating cosine similarity between $q_{model}$ and $n_{model}$.
Specifically, we calculate the positive bank-level similarity $s^{b+}$ as cosine similarity between $\mathbf{q}_{query}$ and $\mathbf{p}_{key}$ , and the negative bank-level similarity $s^{b-}_i$ ($i\in\{1,2,\ldots,K\}$) as cosine similarity between $\mathbf{q}_{query}$ and $i$-th negative sample $\mathbf{n}^i_{key}$ of the memory bank ($K$ is the size of memory bank).

We adopt InfoNCE~\cite{infonce}, a form of contrastive loss
functions, as our objective function for contrastive matching.
The feature-level contrastive loss is formulated as follows:
\begin{equation}
    \mathcal{L}_f = -log\frac{exp(s^{f+}/\gamma)}{exp(s^{f+}/\gamma) + exp(s^{f-}/\gamma)}\text{.}
\end{equation}
The model-level contrastive loss is formulated as follows:
\begin{equation}
    \mathcal{L}_m = -log\frac{exp(s^{m+}/\gamma)}{exp(s^{m+}/\gamma) + exp(s^{m-}/\gamma)}\text{.}
\end{equation}
The bank-level contrastive loss is formulated as follows:
\begin{equation}
    \mathcal{L}_b = -log\frac{exp(s^{b+}/\gamma)}
    {exp(s^{b+}/\gamma) + \sum\limits_{k=1}^{K}exp(s^{b-}_i/\gamma)}\text{,}
\end{equation}
where $K$ is the size of the memory bank and $\gamma$ is a temperature hyper-parameter that is set to be 0.07 in our experiment. Thus, the overall objective function is $\mathcal{L}$:
\begin{equation}
    \mathcal{L} = \alpha_f\mathcal{L}_f + \alpha_m\mathcal{L}_m + \alpha_b\mathcal{L}_b\text{,}
\end{equation}
where $\alpha_f$, $\alpha_m$, and $\alpha_b$ are three hyper-parameters to balance the feature-level, model-level, and bank-level contrasts.

\subsection{Offline Indexing And Retrieval}
After fine-tuning the model on a project-specific dataset, we resort to the offline indexing and retrieval methods proposed by Ciborowska \textit{et. al.}~\cite{ciborowska2022fast}. All encoded changesets are stored in IVFPQ (InVert File with Product Quantization) index. The IVFPQ index is implemented using the Faiss library~\cite{faiss}, which uses the k-means algorithm to partition the embedding space into programmed partitions and assign each embedding to its nearest cluster. In the retrieval process, the query bug report is first located to the nearest partition's centroid, and then the nearest instance within the partition is discovered. For each query bug report, we can identify the $N'$ most similar changesets across all $N$ changesets stored in the IVFPQ index. Therefore, we only re-rank the top-$N'$ subset as the candidate changesets to produce the final ranking.

\section{Experimental Evaluation}
\subsection{Dataset}
The SemanticCodeBERT is trained using all the Java corpus in CodeSearchNet~\cite{codesearchnet}, and we provide the weights and the guidance to fine-tune the pre-trained model for downstream tasks. To evaluate our bug localization technique, we use the dataset separated by Ciborowska \textit{et. al.}~\cite{ciborowska2022fast} from the manually validated dataset by Ming \textit{et. al.}~\cite{bugdataset}. The dataset includes six software projects, termed AspectJ, JDT, PDE, SWT, Tomcat, and ZXing, as shown in Table~\ref{data}. 
To explore the impact of the granularity of changeset data, the bug-inducing changeset is further divided into file-level and hunk-level code changes. Thus, one bug report can have multiple pairs with files
or hunks from the original inducing changes. 
In total, we consider three different granularities: commits, files, and hunks.

\begin{table}
\small
    \centering
    \caption{Six projects used for evaluation.}
    \label{data}
    \resizebox{0.7\linewidth}{!}{
    \begin{tabular}{ccccc}
    \toprule
    \multirow{2}{*}{\textbf{Dataset}} & \multirow{2}{*}{\textbf{Bugs}} & \multicolumn{3}{c}{\textbf{Changesets}} \\
    \cline{3-5} 
    & & \textbf{Commits} & \textbf{Files} & \textbf{Hunks}\\
    \midrule
    AspectJ & 200 & 2,939 & 14,030 & 23,446 \\
    JDT & 94 & 13,860 & 58,619 & 150,630 \\
    PDE & 60 & 9,419 & 42,303 & 100,373 \\
    SWT & 90 & 10,206 & 25,666 & 69,833 \\ 
    Tomcat & 193 & 10,034 & 30,866 & 72,134 \\
    ZXing & 20 & 843 & 2,846 & 6,165 \\
    \bottomrule
    \end{tabular}
    }
\end{table}

\subsection{Evaluation Metrics}
A set of metrics commonly used to evaluate the performance of information retrieval systems are applied to evaluate the performance of different models.

\vspace{0.8mm}
\noindent\textbf{Precision@K ($P@K$)}: $P@K$ evaluates how many of the top-$K$ changesets in a ranking are relevant to the bug report, which is equal to the number of the relevant changesets |$Rel_{B_i}$| located in the top-$K$ position in the ranking averaged across $B$ bug reports:
\begin{equation}
    P@K = \frac{1}{|B|}\sum\limits_{i=1}^{|B|}\frac{|Rel_{B_i}|}{K}\text{.}
\end{equation}
\\
\textbf{Mean Average Precision ($MAP$)}: $MAP$ quantifies the ability of a model to locate all changesets relevant to a bug report. $MAP$ is calculated as the mean of $AvgP$ (average precision) of $B$ bug reports. 

\begin{equation}
    AvgP = \sum\limits_{j=1}^{M}\frac{P@j\times pos(j)}{N}\text{.}
\end{equation}
\begin{equation}
    MAP = \frac{1}{|B|}\sum\limits_{i=1}^{|B|}\frac{1}{AvgP_{B_i}}\text{,}
\end{equation}
where $j$ is the rank, $M$ is the number of retrieved changesets, $pos(j)$ denotes whether $j$-th changeset is relevant to the bug report, $N$ is the total number of bug reports relevant to changesets, $P@j$ is the precision of top-$j$ position in the ranking of this retrieval, and $B_i$ is the $i$-th bug report.

\vspace{0.8mm}
\noindent
\textbf{Mean Reciprocal Rank ($MRR$)}: $MRR$ quantifies the ability of a model to locate the first relevant changeset to a bug report, and is calculated as the average of reciprocal ranks across $B$ bug reports. 1st$Rank_{B_i}$ is the reciprocal rank of $i$-th bug report, which is the inverted rank of the first relevant changeset in the ranking:
\begin{equation}
    MRR = \frac{1}{|B|}\sum\limits_{i=1}^{|B|}\frac{1}{1st Rank_{B_i}}\text{.}
\end{equation}
\\

\vspace{2mm}
\begin{table*}
\footnotesize
    \centering
    \caption{Retrieval performance of different models.}
    \label{SOTA2} 
    \resizebox{\linewidth}{!}
    {
    \begin{tabular}{ll|lllll|lllll|lllll}
    \toprule
    \multirow{2}{*}{\textbf{Projects}} & \multirow{2}{*}{\textbf{Technique}} & \multicolumn{5}{c}{\textbf{$Commits-$}} & \multicolumn{5}{c}{\textbf{$Files-$}} & \multicolumn{5}{c}{\textbf{$Hunks-$}} \\
    \cline{3-17}
    &  & \textbf{MRR} & \textbf{MAP} & \textbf{P@1} & \textbf{P@3} & \textbf{P@5} & \textbf{MRR} & \textbf{MAP} & \textbf{P@1} & \textbf{P@3} & \textbf{P@5} & \textbf{MRR} & \textbf{MAP} & \textbf{P@1} & \textbf{P@3} & \textbf{P@5}\\
    \midrule
    \multirow{5}{*}{ZXing} 
    & \textit{BLUiR} & 0.077 & 0.016 & 0.071 & 0.024 & 0.014 & 0.073 & 0.023 & 0.000 & 0.024 & 0.014 & 0.056 & 0.035 & 0.000 & 0.071 & 0.086 \\

    & \textit{FBL-BERT} &0.155 &0.061& 0.100 &0.133& 0.120 &0.212 &0.163& 0.100 &0.133& 0.220  &0.328 &0.210& 0.200 &0.233& 0.240  \\
    
    & \textit{GraphCodeBERT} & 0.189 &0.118& 0.143 &0.143& 0.118 & 0.280 &0.155& 0.214 &0.143& 0.214 &0.346 &0.118& 0.225 &0.111& 0.067\\

    & \textit{UniXcoder} & 0.354 & 0.167 & 0.414 & 0.171 & 0.120 & 0.359 & 0.143 & 0.333 & 0.224 & 0.200 & 0.331 & 0.164 & 0.214 & 0.261 & 0.282 \\
    & \textit{Ours} & \bf{0.439} &\bf{0.226}& \bf{0.429} &\bf{0.250}&\bf{0.225} &\bf{0.421}  & \bf{0.185} & \bf{0.357}  & \bf{0.226}  & \bf{0.271} & \bf{0.422} & \bf{0.212} & \bf{0.333} &\bf{0.444} & \bf{0.400} \\

    \midrule
     \multirow{5}{*}{PDE} 
     & \textit{BLUiR} & 0.009 & 0.001 & 0.000 & 0.000 & 0.000 & 0.018 & 0.003 & 0.000 & 0.008 & 0.005 & 0.024 & 0.005 & 0.000 & 0.008 & 0.010 \\

     & \textit{FBL-BERT} &0.103 &0.013& 0.067 &0.033& 0.027 &0.260 &0.079& 0.167 &0.128& 0.151 &0.288 &0.093& 0.200 &0.144& 0.127 \\
   
    & \textit{GraphCodeBERT} & 0.180 & 0.042 & 0.142 & 0.087 & 0.058 & 0.264 &0.094 & 0.167 &0.129 & 0.148  & 0.284 &0.074& 0.206 &0.124& 0.129 \\
    
    & \textit{UniXcoder} & 0.178 & 0.029 & 0.095 & 0.063 & 0.072 & 0.267 & 0.090 & 0.167 & 0.135 & 0.129 & 0.289 & 0.102 & 0.212 & 0.144 & 0.129 \\

    & \textit{Ours} &\bf{0.248} &\bf{0.045}& \bf{0.190} &\bf{0.103}& \bf{0.076} & \bf{0.274} &\bf{0.095}& \bf{0.214} &\bf{0.137}& \bf{0.160} & \bf{0.294} &\bf{0.134}& \bf{0.286} &\bf{0.182}& \bf{0.160}\\

    \midrule
     \multirow{5}{*}{AspectJ} 
     & \textit{BLUiR} & 0.016 & 0.013 & 0.007 & 0.014 & 0.015 & 0.098 & 0.065 & 0.028 & 0.076 & 0.108 & 0.086 & 0.048 & 0.007 & 0.017 & 0.159 \\

     & \textit{FBL-BERT} & 0.107 &0.061& 0.058 &0.080& 0.083 &0.176 &0.085& 0.154 &0.095& 0.097 &0.183 &0.093& 0.173 &0.111& 0.099  \\

    & \textit{GraphCodeBERT} & 0.172 &0.065& 0.167 &0.065& 0.060 & 0.178 &0.071& 0.167 &0.065& 0.060 &  0.188 &0.086 & 0.167 & 0.120 & 0.116  \\
    & \textit{UniXcoder} & 0.270 & 0.148 & 0.245 & 0.160 & 0.158  & 0.209 & 0.119 & 0.167 & 0.140 & \bf{0.152} & 0.250 & 0.134 & 0.250 & 0.150 & 0.138 \\

    & \textit{Ours} & \bf{0.309}  &\bf{0.169} & \bf{0.278}  &\bf{0.198} & \bf{0.196}   &\bf{0.272}&\bf{0.148}& \bf{0.250} & \bf{0.157}& 0.146 & \bf{0.262} &\bf{0.143}& \bf{0.250} &\bf{0.161}& \bf{0.163} \\

    \midrule
     \multirow{5}{*}{JDT} 
     & \textit{BLUiR} & 0.019 & 0.001 & 0.015 & 0.005 & 0.003 & 0.027 & 0.003 & 0.000 & 0.010 & 0.012 & 0.033 & 0.005 & 0.000 & 0.005 & 0.009 \\

     & \textit{FBL-BERT} &0.118 &0.016& 0.064 &0.043& 0.030 &0.403 &0.060& 0.319 &0.184& 0.128  & 0.429 &0.062& 0.319 &0.195& 0.167 \\

    & \textit{GraphCodeBERT} & 0.125 &0.022& 0.061 &0.035& 0.030 & 0.423 &0.058& 0.308 &0.179& 0.118  &0.385 &0.041& 0.231 &0.179& 0.118 \\

    & \textit{UniXcoder} & 0.182 & 0.018 & 0.182 & 0.061 & 0.038 & 0.434 & 0.062 & 0.379 & 0.166 & 0.131 & 0.364 & 0.045 & 0.288 & 0.182 & 0.123 \\

    & \textit{Ours} & \bf{0.306} &\bf{0.026} & \bf{0.288}  &\bf{0.096} & \bf{0.064}  & \bf{0.489} &\bf{0.080}& \bf{0.462} & \bf{0.195} & \bf{0.167} & \bf{0.443} &\bf{0.088} & \bf{0.322} &\bf{0.206} & \bf{0.167} \\

    \midrule
     \multirow{5}{*}{SWT}
    & \textit{BLUiR} & 0.005 & 0.001 & 0.000 & 0.000 & 0.000 & 0.020 & 0.003 & 0.016 & 0.005 & 0.006 & 0.014 & 0.001 & 0.000 & 0.000 & 0.013 \\

     & \textit{FBL-BERT} &0.067 &0.015& 0.023 &0.027& 0.026 & 0.555 &0.131 &0.535 &0.233 &0.173&0.526 & 0.131 & 0.488 & 0.217 & 0.164 \\

    & \textit{GraphCodeBERT} & 0.105 &0.018& 0.048 &0.026& 0.022  & 0.535 &0.137& 0.525 &0.220 & 0.175 & 0.536 &0.132 & 0.516 & 0.220 & 0.159 \\

    & \textit{UniXcoder} & 0.129 & 0.035 & 0.107 & 0.106 & 0.063 & 0.548 & 0.149 & 0.524 & 0.233 & 0.183 & 0.535 & 0.143 & 0.535 & 0.205 & 0.179 \\

    & \textit{Ours} & \bf{0.283}  &\bf{0.085} & \bf{0.159}  &\bf{0.177} & \bf{0.170}  &\textbf{0.560} &\textbf{0.153}& \textbf{0.540} &\textbf{0.249}& \textbf{0.192}  & \textbf{0.540} &\textbf{0.147}& \textbf{0.540} &\textbf{0.228}& \textbf{0.179}\\

    \midrule
     \multirow{5}{*}{Tomcat} 
    & \textit{BLUiR} & 0.007 & 0.002 & 0.000 & 0.002 & 0.002 & 0.014 & 0.003 & 0.000 & 0.010 & 0.007 & 0.014 & 0.005 & 0.000 & 0.012 & 0.013 \\

     & \textit{FBL-BERT} & 0.141 &0.055& 0.062 &0.077& 0.088 & 0.463 & 0.114& 0.381 &0.222& 0.183 & 0.482 & 0.129 & 0.412 & 0.216 & 0.182 \\
   
    & \textit{GraphCodeBERT} & 0.253 &0.062& 0.188 &0.104& 0.084 & 0.287 &0.067& 0.271 &0.104& 0.080  & 0.395 &0.118& 0.363 &0.216& 0.211 \\
    & \textit{UniXcoder} & 0.328 & 0.057 & 0.338 & 0.120 & 0.084 & 0.364 & 0.065 & 0.353 & 0.125 & 0.085 & 0.396 & 0.097 & 0.378 & 0.139 & 0.118 \\

    & \textit{Ours} &\textbf{0.386}& \textbf{0.073} & \textbf{0.360}  & \textbf{0.135}& \textbf{0.107}  & \textbf{0.487} &\textbf{0.122}& \textbf{0.406} &\textbf{0.247}& \textbf{0.232} & \textbf{0.484} & \textbf{0.132} & \textbf{0.423} & \textbf{0.225} & \textbf{0.211} \\

    \bottomrule
    \end{tabular}
    }
\end{table*}

\subsection{Experimental Setup}

\textbf{Configurations of Pre-training Tasks}: 
The SemanticCodeBERT is pre-trained on NVIDIA Tesla A100 with 128GB RAM on the Ubuntu system. The Adam optimizer~\cite{adam} is used to update model parameters with batch size 80 and learning rate 1E-04. To accelerate the training process, the parameters of GraphCodeBERT~\cite{graphcodebert} are used to initialize the pre-training model. The model is trained with 600K batches and costs about 156 hours.

\vspace{0.8mm}
\noindent
\textbf{Configurations of Bug Localization}:
The first half of the project’s pairs of bug reports and bug-inducing changesets, ordered by bug opening date, are selected as the training dataset, and the remaining half is left as the test dataset. 
The experiments are implemented with GPU support. The Adam optimizer~\cite{adam} is used to update model parameters with learning rate 3E-05. 
All bug reports and changesets are truncated or padded to their respective length limit. According to the experimental verification, we set the trade-off hyper-parameters $\alpha_f$, $\alpha_m$, and $\alpha_b$ as $1$, $1$, and $1$, respectively.

\vspace{0.8mm}
\noindent
\textbf{Changeset Encoding Strategies}:
Changesets are time-ordered sequences recording the software's evolution over time. We build upon the three changeset encoding strategies ($D$-encoding, $ARC$-encoding, and $ARC_L$-encoding) proposed by Ciborowska \textit{et. al.}~\cite{ciborowska2022fast} to encode changesets.
$D$-encoding does not utilize specific characteristics of changeset lines. $ARC$-encoding divides the lines into three groups with three unique tokens. $ARC_L$-encoding instead does not group the lines and maintains the ordering of lines within a changeset. 
These three strategies are based on the output of the git \textit{diff} command, which divides changeset lines into 
three kinds: added lines, removed lines, and unchanged lines. All code sequences are preprocessed by filtering the intrusive characters (\textit{e.g.,} docstrings, comments) from the original code tokens.

\subsection{Retrieval Performance}
We compare the performance of our proposed model with the traditional bug localization tool, state-of-the-art changeset-based bug localization approach, and two recent state-of-the-art pre-trained models with the HMCBL framework. 
\begin{itemize}[leftmargin=*]
    \item \textbf{BLUiR}~\cite{saha2013improving}: A structured IR-based fault localization tool, which builds AST to extract the program constructs of each source code file and utilizes Okapi BM25~\cite{Okapi} to calculate the similarity between the bug report and the candidate changesets. 
    \vspace{0.2mm}
    \item \textbf{FBL-BERT}~\cite{ciborowska2022fast}: The state-of-the-art approach for automatically retrieving bug-inducing changesets given a bug report, which uses the popular BERT model to more accurately match the semantics in the bug report text with the bug-inducing changesets.
    
    \vspace{0.2mm}
    \item \textbf{GraphCodeBERT}~\cite{graphcodebert}: A pre-trained model that considers data flow to better encode the relation between variables. 
    \vspace{0.2mm}
    \item \textbf{UniXcoder}~\cite{unixcoder}: An unified cross-modal pre-trained model, which leverages cross-modal information like Abstract Syntax Tree and comments to enhance code representation. 
\end{itemize}

For BLUiR, we fully follow the original technical description in~\cite{saha2013improving} (as no open-source implementation is available) to get the results for the evaluation metrics. 
For FBL-BERT, we use the experimental results provided in~\cite{ciborowska2022fast}. 
For GraphCodeBERT and UniXcoder, we get the results by replacing the pre-trained model SemanticCodeBERT within the HMCBL framework respectively with GraphCodeBERT and UniXcoder (keeping other configurations the same).
Table~\ref{SOTA2} shows the retrieval performances of different models with different changeset encoding strategies (\textit{i.e.,} $D$-, $ARC$- and $ARC_L$- encoding) and three granularities (\textit{i.e.} $Commits-$, $Files-$ and $Hunks-$ level) on six projects.
Limited by space, the best result of the three encoding strategies is shown for each configuration. 

The following observations can be obtained from the figure. First, compared with the traditional bug localization method which relies on more direct term matching between a bug report and a changeset, the neural network methods perform better by obtaining semantic representations for the calculation of similarity. 
Second, our proposed method outperforms the state-of-the-art method (FBL-BERT) by a clear margin. In particular, our proposed bug localization technique improves FBL-BERT by 140.78\% to 188.79\% in terms of MRR on six projects with $Commits-$ level granularity. 
Third, compared with GraphCodeBERT and UniXcoder, our model using SemanticCodeBERT as a changeset encoder consistently achieves better performance in almost all experimental configurations. This suggests that the proposed Semantic Flow Graph (SFG) captures good code semantics, and the proposed framework contributes to changeset-based bug localization.

The Student’s t-test is conducted between our technique and other
baselines, and the results show that the improvements are significant with p < 0.01.
We additionally observe that with the $Commits-$level granularity, the obtained improvement is more significant than the other two granularities ($Files-$level and $Hunks-$ level). It can be attributed that the undivided bug-inducing changeset carries enriched semantic information which can be captured by SemanticCodeBERT. 
This again confirms the effectiveness of the SemanticCodeBERT-based bug localization technique.

\begin{table}
\footnotesize
    \centering
    \caption{Ablation study of pre-training tasks of SemanticCodeBERT with Semantic Flow Graph (SFG).}
    \label{ablation1}
    \resizebox{\linewidth}{!}{
    \begin{tabular}{lllllll}
    \toprule
    \textbf{Dataset} & \textbf{Pre-training Tasks} & \textbf{MRR} & \textbf{MAP} & \textbf{P@1} & \textbf{P@3} & \textbf{P@5} \\
\midrule
    \multirow{3}{*}{ZXing} & -w/ & 0.189 &0.118& 0.143 &0.143& 0.118 \\
    &-w/ N.\& E. & 0.372 & 0.102 & 0.333 & 0.111 & 0.067 \\
    &-w/ N.\& E.\& T.\& R.  & \textbf{0.439}& \textbf{0.226} & \textbf{0.429} & \textbf{0.250} & \textbf{0.225} \\
\midrule
    \multirow{3}{*}{PDE} & -w/ & 0.180 & 0.042 & 0.142 & 0.087 & 0.058 \\
    &-w/ N.\& E. & 0.219 & 0.032 & 0.143 & 0.076 & 0.072 \\
    &-w/ N.\& E.\& T.\& R.  & \textbf{0.248} & \textbf{0.045} & \textbf{0.190} & \textbf{0.103} & \textbf{0.076} \\
\midrule
    \multirow{3}{*}{AspectJ} & -w/ & 0.172& 0.065 & 0.167 & 0.065 & 0.060 \\
    &-w/ N.\& E. & 0.289 & 0.158 & 0.250 & 0.184 & 0.170 \\
    &-w/ N.\& E.\& T.\& R.  & \textbf{0.309}& \textbf{0.169} & \textbf{0.278} & \textbf{0.198} & \textbf{0.196} \\

\midrule
    \multirow{3}{*}{JDT} & -w/ & 0.125 &0.022& 0.061 &0.035& 0.030 \\
    &-w/ N.\& E. & 0.139 & 0.021 & 0.095 & 0.044 & 0.048 \\
    &-w/ N.\& E.\& T.\& R.  & \textbf{0.306} & \textbf{0.026} & \textbf{0.288} & \textbf{0.096} & \textbf{0.064} \\
\midrule
    \multirow{3}{*}{SWT} & -w/ & 0.105 & 0.018 & 0.048 & 0.026 & 0.022 \\
    &-w/ N.\& E. & 0.197 & 0.058 & 0.063 & 0.085 & 0.141 \\
    &-w/ N.\& E.\& T.\& R.  & \textbf{0.283} & \textbf{0.085} & \textbf{0.159} & \textbf{0.177} & \textbf{0.170} \\

\midrule
    \multirow{3}{*}{Tomcat} & -w/ & 0.253 & 0.062 & 0.188 & 0.104 & 0.084 \\
    &-w/ N.\& E. & 0.300 & 0.048 & 0.346 & 0.113 & 0.077 \\
    &-w/ N.\& E.\& T.\& R. & \textbf{0.386} & \textbf{0.073} & \textbf{0.360}  & \textbf{0.135} & \textbf{0.107} \\

    \bottomrule
    \end{tabular}
    }
\end{table}

\vspace{2mm}
\begin{table}
\footnotesize
    \centering
    \caption{Ablation study of Hierarchical Momentum Contrastive Bug Localization (HMCBL) technique, where GCBERT and SCBERT are short of GraphCodeBERT and SemanticCodeBERT.}
    \label{ablation2}
    \resizebox{\linewidth}{!}{
    \begin{tabular}{lllllll}
    \toprule
    \textbf{Technique} & \textbf{Dataset} & \textbf{MRR} & \textbf{MAP} & \textbf{P@1} & \textbf{P@3} & \textbf{P@5} \\
    \midrule
    \multirowcell{6}[0pt][l]{\textit{BERT} -w/o \\ \textit{HMCBL} \\ (\textit{FBL-BERT})}
    & ZXing &0.155 &0.061& 0.100 &\textbf{0.133}& \textbf{0.120} \\
    & PDE  &0.103 &0.013& 0.067 &0.033& 0.027 \\
    & AspectJ & 0.107 &0.061& 0.058 &0.080& 0.083 \\
    & JDT & 0.118 & 0.016 & \textbf{0.064} & 0.043 & 0.030 \\
    & SWT &0.067 &\textbf{0.015}& 0.023 &\textbf{0.027}& \textbf{0.026} \\
    & Tomcat & 0.141 &0.055& 0.062 &0.077& 0.088 \\

    \midrule
    \multirowcell{6}[0pt][l]{\textit{GCBERT} -w/o \\ \textit{HMCBL}}
    & ZXing & 0.162 & 0.106 & 0.143 & 0.095 & 0.086 \\
    & PDE  & 0.167 & 0.018 & 0.119 &0.071& 0.045 \\
    & AspectJ & 0.123 & 0.067 & 0.076 & \textbf{0.073} & \textbf{0.084} \\
    & JDT & 0.120 & 0.022 & 0.061 & 0.035 & \textbf{0.036} \\
    & SWT & 0.090 & \textbf{0.019} & 0.048 & 0.021 & 0.022 \\
    & Tomcat & 0.151 & 0.035 & 0.059 & 0.064 & 0.063 \\
    
    \midrule
    \multirowcell{6}[0pt][l]{\textit{SCBERT} -w/o \\ \textit{HMCBL}}
    & ZXing & 0.222 &0.112& 0.143 &0.190 & 0.150 \\
    & PDE & 0.230  & \textbf{0.049} & 0.142  & 0.095 & 0.069 \\
    & AspectJ & 0.271 & 0.148 & 0.250  & 0.161 & 0.165 \\
    & JDT & 0.217 & \textbf{0.051} & 0.136 & \textbf{0.111} & \textbf{0.091} \\
    & SWT & 0.250 & 0.062 & 0.095  & 0.167 & \textbf{0.185} \\
    & Tomcat & 0.285 & 0.053 & 0.265 & 0.092 & 0.069 \\

    \hline 
    \midrule

    \multirowcell{6}[0pt][l]{\textit{BERT} -w/ \\ \textit{HMCBL} \\ }
    & ZXing & \textbf{0.179} & \textbf{0.040} & \textbf{0.143} & 0.095 & 0.061 \\
    & PDE & \textbf{0.156} & \textbf{0.032} & \textbf{0.119} & \textbf{0.063} & \textbf{0.051} \\
    & AspectJ & \textbf{0.162}& \textbf{0.097} & \textbf{0.118} & \textbf{0.141} & \textbf{0.149} \\
    & JDT & \textbf{0.128} & \textbf{0.017} & 0.030 & \textbf{0.070} & \textbf{0.100} \\
    & SWT & \textbf{0.082 }& 0.013 & \textbf{0.048} & 0.024 & 0.021 \\
    & Tomcat & \textbf{0.235} & \textbf{0.055} & \textbf{0.169} & \textbf{0.098} & \textbf{0.096} \\

    \midrule
    \multirowcell{6}[0pt][l]{\textit{GCBERT} -w/ \\ \textit{HMCBL}}
    & ZXing & \textbf{0.189} &\textbf{0.118}& \textbf{0.143} &\textbf{0.143}& \textbf{0.118} \\
    & PDE  & \textbf{0.180} & \textbf{0.042} & \textbf{0.142} & \textbf{0.087} & \textbf{0.058} \\
    & AspectJ & \textbf{0.172}& \textbf{0.065} & \textbf{0.167} & 0.065 & 0.060 \\
    & JDT & \textbf{0.125} &\textbf{0.022}& \textbf{0.061} &\textbf{0.035}& 0.030\\
    & SWT & \textbf{0.105} & 0.018 & \textbf{0.048} & \textbf{0.026} & \textbf{0.022} \\
    & Tomcat & \textbf{0.253} & \textbf{0.062} & \textbf{0.188} & \textbf{0.104} & \textbf{0.084} \\

    \midrule
    \multirowcell{6}[0pt][l]{\textit{SCBERT} -w/ \\ \textit{HMCBL}}
    & ZXing & \textbf{0.439} & \textbf{0.226} & \textbf{0.429} &\textbf{0.250} & \textbf{0.225} \\
    & PDE  & \textbf{0.248} & 0.045 & \textbf{0.190} & \textbf{0.103} & \textbf{0.076} \\
    & AspectJ & \textbf{0.309}& \textbf{0.169} & \textbf{0.278} & \textbf{0.198} & \textbf{0.196} \\
    & JDT & \textbf{0.306} & 0.026 & \textbf{0.288} & 0.096 & 0.064 \\
    & SWT & \textbf{0.283} & \textbf{0.085} & \textbf{0.159} & \textbf{0.177} & 0.170 \\
    & Tomcat & \textbf{0.386} & \textbf{0.073} & \textbf{0.360}  & \textbf{0.135} & \textbf{0.107} \\
    
    \bottomrule
    \end{tabular}
    }
\end{table}

\subsection{Ablation Study}
To evaluate the design choices in the proposed model, we conduct several ablation studies. 
To begin with, as shown in Table~\ref{ablation1}, we analyze the contributions of node alignment, edge prediction, type prediction, and role prediction pre-training tasks on the six projects with commits granularity. N., E., T., and R. denote the Node Alignment, Edge Prediction, Type Prediction, and Role Prediction pre-training tasks, respectively. With all of these pre-training tasks, we train SemanticCodeBert according to the proposed new code representation SFG. According to the results, after adding Type and Role Prediction pre-training tasks, the obtained performance has universally improved. This result suggests that leveraging the node attributes (type and role) is vital to learn code representation. 

Furthermore, we evaluate the effectiveness of the Hierarchical Momentum Contrastive Bug Localization (HMCBL) technique on the six projects with commits granularity. As illustrated in Table~\ref{ablation2}, for -w/o HMCBL, the memory bank and hierarchical contrastive loss which leverages similarities at different levels do not exist, and only the representation obtained by the encoder is utilized to calculate similarity.

To demonstrate the generality, the technique is evaluated with different pre-training models as the encoder of the changeset, including BERT, GraphCodeBERT, and SemanticCodeBERT. It is observed that overall much better performance will be obtained with hierarchical momentum contrastive learning, which provides large-scale negative sample interactions for representation learning and increases retrieval accuracy. For instance, compared with BERT -w/o HMCBL, which is the FBL-BERT exactly, BERT -w/ HMCBL improves the performance in terms of MRR scores for more than 80\% projects by 15.48\% to 66.67\%.
It is indicative of the observation that the hierarchical momentum contrastive bug localization technique can be extended as a general and effective framework with different advanced pre-training models.

\subsection{Threats to Validity} 
Our results should be interpreted with several threats to validity in mind. As bug-inducing changes are identified using the SZZ algorithm~\cite{SZZ}, one threat to the internal validity of the results is possible noise introduced by SZZ may make the mapping between bug reports and bug-inducing changesets not very precise. However, the dataset used in the study has been validated manually~\cite{bugdataset}, so this threat is minimized. Another threat to internal validity is the dataset may contain tangled changes~\cite{tanglechange}. While we do believe tangled changes can affect our results, the dataset has been widely used for changeset-based bug localization studies~\cite{bugdataset,ciborowska2022fast}, and removing tangled changes completely is extraordinarily difficult.

With regard to threats to external validity, one potential issue is that the evaluation is conducted on a limited number of bugs from several open-source projects. However, these projects feature various purposes and development styles. Also, the dataset can be considered as the de-facto evaluation target for changeset-based bug localization studies and prior studies have widely used it~\cite{bugdataset,ciborowska2022fast}.

\section{Conclusion}

We aim to advance the state-of-the-art BERT-based bug localization techniques in this paper, which currently suffer from two issues: the pre-trained BERT models on source code are not robust enough to capture code semantics and the overall bug localization models neglect the necessity of large-scale negative samples in contrastive learning and ignore the lexical similarity between bug reports and changesets. To address these two issues, we 1) propose a novel directed, multiple-label Semantic Flow Graph (SFG), which
compactly and adequately captures code semantics, 2) design and train SemanticCodeBERT on the basis of SFG, and 3) design a novel Hierarchical Momentum Contrastive Bug Localization technique (HMCBL). Evaluation results confirm that our method achieves state-of-the-art performance. 

\section{Data Availability}
Our replication package (including code, model, etc.) is publicly available at \url{https://github.com/duyali2000/SemanticFlowGraph}.

\section*{Acknowledgments}
\noindent
We are grateful to the anonymous ESEC/FSE reviewers for their valuable feedback on this work. This work was partially supported by National Natural Science Foundation of China (Grant No. 62102233), Shandong Province Overseas Outstanding Youth Fund (Grant No. 2022HWYQ-043), and Qilu Young Scholar Program of Shandong University. 

\bibliographystyle{ACM-Reference-Format}
\balance
\bibliography{sample-base}

\end{document}